\documentclass[11pt, letter, a4paper]{article}
\usepackage[it, bf]{caption}
\usepackage{graphics}
\usepackage{graphicx}
\usepackage{epsfig}
\usepackage{amssymb}
\usepackage{amsmath}
\usepackage{txfonts}
\begin{document}
\setcounter{page}{152}
\begin{center}
\Large{\textbf{On the coupled hydraulic and dielectric material properties of soils: combined numerical and experimental investigations\footnote{Proc. 9th International Conference on Electromagnetic Wave Interaction with Water and Moist Substances, ISEMA 2011, Editor: David B. Funk,
Kansas City, Missouri USA, May 31 - June 3, 152-161, 2011}}}\vspace{0.5cm}\\
 \large{\textmd{Norman Wagner$^{(1)}$, Alexander Scheuermann$^{(2,3)}$, Moritz Schwing$^{(2)}$, \\ Frank Bonitz$^{(1)}$, Klaus Kupfer$^{(1)}$}}\vspace{0.2cm}\\
{ \normalsize{{\textmd{$^{(1)}$ Institute of Material Research and Testing at the Bauhaus-University Weimar\\Coudraystr. 4, 99423 Weimar, Germany, Email: {norman.wagner@mfpa.de}\\}}}
 \normalsize{{\textmd{$^{(2)}$ The Golder Geomechanics Centre, University of Queensland, Brisbane, Australia\\}}}
 \normalsize{{\textmd{$^{(3)}$ Institute for Soil Mechanics and Rock Mechanics, University of Karlsruhe, Germany\\
 Karlsruhe Institute of Technology (KIT)}
 \vspace{0.05cm}}}
}
 \end{center}
\subsection*{\textit{ABSTRACT}}
\textit{Precise knowledge of the frequency dependent electromagnetic properties of porous media is urgently necessary for successful utilization of high frequency electromagnetic measurement techniques for near and subsurface sensing. Thus, there is a need of systematic investigations by means of dielectric spectroscopy of unsaturated and saturated soils under controlled hydraulic conditions. In this context, two-port rod based transmission lines (R-TMLs) were characterized in the frequency range from 1~MHz to 10~GHz by combined theoretical, numerical, and experimental investigations. To analyze coupled hydraulic and dielectric soil properties a slightly plastic clay soil was investigated. There is evidence that the bound water contribution of the soil is substantially lower than expected.
}\vspace{0.2cm}

\noindent{\it Keywords: soil matric potential, soil moisture, dielectric relaxation behavior of soil}
\section{\normalsize{{INTRODUCTION}}}\label{sec:intro}

Despite the fact that a precise knowledge of the frequency dependent electromagnetic material properties of partially saturated porous media, e.g. soil, is urgently necessary for a successful application of high frequency (radio and microwave) electromagnetic (HF-EM) moisture measurement techniques (Time and Frequency Domain Reflectometry, active and passive Remote Sensing, Ground Penetration Radar, capacitive methods) for near and subsurface sensing there is a lack of temperature and frequency dependent HF-EM soil properties as well as appropriate accurate broadband mixture approaches \cite{Robi03, Chen2006, Arcone2008, Kupf07, Schwartz2009, Wagner2011}.

In particular, in fine-grained soils the movement of water is influenced by different surface bonding forces due to interface processes \cite{Iwata1995}. The interface effects lead to a number of dielectric relaxation processes (free, interface or confined water phase, Maxwell-Wagner-effect, counterion relaxation effects \cite{Ishi00, Ishi03, Logs04, Chen2006}). These relaxation processes are the reason for a strong frequency dependence of the HF-EM material properties below 1 GHz \cite{Asan07, Arcone2008, Wagn07, Wagner2011} effecting HF-EM techniques. Hence, there is a need of systematic investigations by broadband dielectric spectroscopy of unsaturated and saturated soils under controlled hydraulic and mechanical conditions. In this context, two-port rod based transmission lines (R-TMLs) were characterized in the frequency range from 1~MHz to 10~GHz using combined theoretical, numerical and experimental investigations. The propagation characteristics and sensitivity of the R-TMLs was determined by 3D numerical finite element calculations according to \cite{Wagn07a} and measurements on standard materials as well as nearly saturated and unsaturated soils. The frequency dependent complex permittivity was determined by means of analytical or numerical inversion of measured four complex S-parameters. The results were compared with broadband coaxial transmission line (C-TML) and open ended coaxial line (O-TML) techniques (see \cite{Wagner2011}). Furthermore, a five rod R-TML setup was developed on N-based connectors.

\section{\normalsize{COUPLING HYDRAULIC AND DIELECTRIC SOIL PROPERTIES}}\label{sec:model}

Organic free soil as a typical porous medium consists mainly of four phases: solid particles (various
mineral phases), pore air, pore fluid as well as a solid particle - pore fluid interface.
In principle the fractions of the soil phases vary both
in space (due to composition and density of the soil) and time (due to changes of water content,
porosity, pore water chemistry and temperature). The electromagnetic properties of the solid particles are frequency
independent in the considered temperature-pressure-frequency range.
Real relative permittivity varies from 3 to 15 \cite{Robinson2004}.
The pore fluid as well as interface fluid are mainly an aqueous solution with a temperature-pressure-frequency dependent relative
complex permittivity $\varepsilon^\star_{w} (\omega ,T, p)$
according to the modified Debye model \cite{Kaat07}.
Under atmospheric conditions the dielectric relaxation time of water $\tau_w(T)$ depends on temperature
$T$ according to the Eyring equation \cite{WagScheu2009a} with Gibbs energy or free enthalpy of activation
$\Delta G_w^\ddag(T) = \Delta H_w^\ddag (T) -T\Delta
S_w^\ddag (T)$, activation enthalpy $\Delta H_w^\ddag(T)$ and
activation entropy $\Delta S_w^\ddag(T)$. Furthermore, Gibbs energy of the interface fluid $\Delta G_d^\ddag(T)$
is assumed to be a function of the distance from the particle surface (for quantitative approaches see \cite{WagScheu2009a}).

Soil matric potential $\Psi_m$ is a measure of the bonding forces on water in the soil and is related to the
chemical potential of soil water $\Delta\mu_W = \mu_W^\circ - \mu_W = \Psi_m V_w$
with chemical potential at a reference state $\mu_W^\circ$ and molar volume of water $V_W$ \cite{Iwata1995, Hilh01}.
Thus, Hilhorst et al. \cite{Hilhorst1998, Hilh01, Hilhorst2000} suggested an
approach for the relationship between $\Psi_m$ and $\Delta G_d^\ddag$:
\begin{equation}\label{eq:matric1}
\Psi_m(T)\cdot V_W = \Delta G_{w}^{\ddag\circ}(T) - \Delta G_d^\ddag(T)
\end{equation}
with Gibbs energy of water at a reference state $\Delta G_{w}^{\ddag\circ}(T)$
($10.4$\hspace{0.1cm}kJ/mol at atmospheric conditions and $T$=293.15 K).
This relationship is used to calculate Gibbs energy or free enthalpy of dielectric activation of interface water
$\Delta G_d^\ddag(T)$. The complex relative dielectric permittivity of free and interface water of a
porous material, e.g. soil, in dependence of the volumetric water content $\theta$ under atmospheric conditions then can
be calculated \cite{Wagn09}:
\begin{equation}\label{eq:matric2}
 \varepsilon_{a(\theta, n)}^\star (\theta ,T) = \int\limits_{\Psi_m(0)}^{\Psi_m(\theta)}  {\varepsilon^{\star a(\theta, n)}_w(\Psi_m(\theta)
 ,T)}\frac{d\theta(\Psi_m)}{d\Psi_m}d\Psi_m.
\end{equation}
The parameter $0\leq a \leq 1$ is defined by the used mixture approach to
obtain the effective complex relative permittivity of the soil $\varepsilon^\star_{r,\mbox{eff}} (\theta,T )$.
The parameter $a$ contains in principle structural information of free and interface water in the porous material and is strictly speaking
a function of volumetric water content $\theta$ and porosity $n$.
However, Hilhorst 1998 (HIL, \cite{Hilhorst2000}) suggests the following equation:
\begin{equation}\label{eq:HilMix}
 \varepsilon^\star_{r,\mbox{eff}} (\theta,T ) =
 \frac{\varepsilon_{1}^\star (\theta,T )}{3(2n - \theta )}  + (1 - n)\varepsilon _G(T) + (n - \theta ).
\end{equation}
with porosity $n$ and real relative permittivity of solid grain $\varepsilon _G$. As an alternative approach
Wagner et al. 2011 \cite{Wagner2011} suggests the so called advanced Lichtenecker and Rother Model (ALRM):
\begin{equation}\label{eq:LRMix}
\varepsilon_{r,\mbox{eff}}^{\star a(\theta, n)} (\theta,T ) = \varepsilon_{a(\theta, n)}^\star(\theta,T ) + (1 - n)\varepsilon
_G(T)^{a(\theta, n)} + (n - \theta )
\end{equation}
which is frequently used in CRIM form with a constant structure parameter $a=0.5$ \cite{Mironov2004, WagScheu2009a}.
Furthermore, according to \cite{Wagner2011} the following empirical relationships for the structure parameter $a$ and
pore water conductivity $\sigma_W^\circ$ with calibration constants $A_i$, $B_i$, $C_i$ were used:
\begin{equation}\label{eq:LichtRothM2}
 a (\theta ,n) = A_{1}  + B_{1} n^2  + C_{1} \left(\frac{\theta}{n}\right)^2
\end{equation}
as well as
\begin{equation}\label{eq:LichtRothM3}
\log \left( {\sigma^\circ_{W} (\theta ,n)} \right) = \log A_{2}  + B_{2} n + C_{2}  \left(\frac{\theta}{n}\right).
\end{equation}

\section{\normalsize{SYSTEMATIC ANALYSIS OF ROD BASED TRANSMISSION LINE CELLS}}\label{sec:Soil}
For the determination of the propagation characteristics and sensitivity of the R-TMLs 3D finite element calculations in Ansoft-HFSS (High Frequency Structure Simulator) as well as vector network analyzer measurements (Rhode  \& Schwarz ZVR, 1~MHz - 4 GHz, Agilent PNA, 10~MHz - {10~GHz}) were performed: (i) on different rod numbers, thicknesses as well as arrangements in air, (ii) with electrical and dielectric low loss and high loss strong dispersive standard materials (liquids, saturated and dry glass, zircon and baddeleyite beads) and (iii) on saturated and unsaturated soil.

\subsection{Numerical 3D HF-EM calculations}
\begin{figure}[h]
 \center
  \includegraphics[scale=0.4]{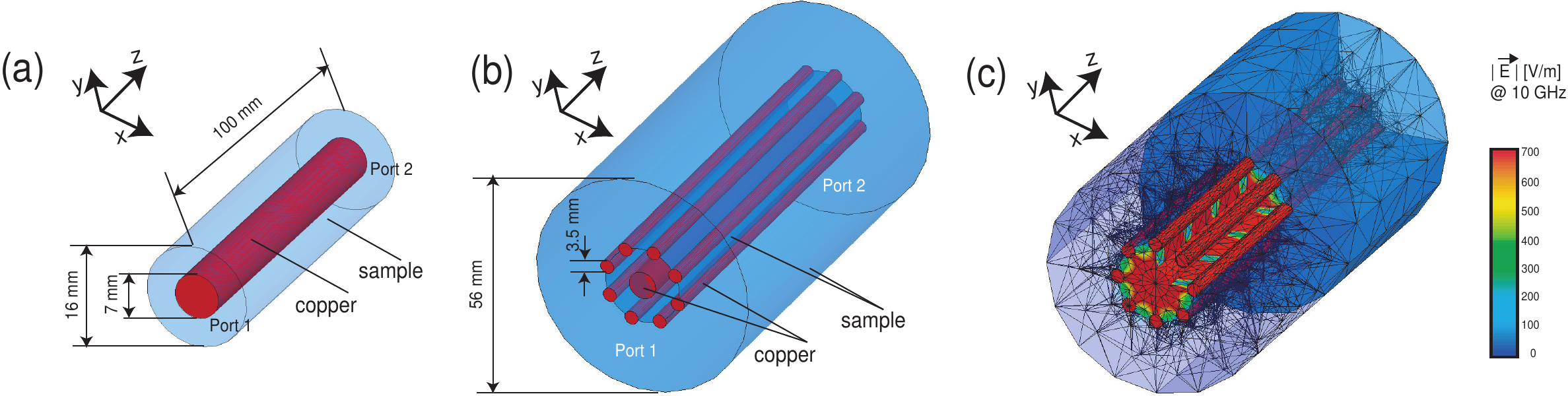}\\
  \caption{\textit{Model geometry of the transmission line cell: (a) conventional coaxial cell and (b) rod based cell. (c) Tetrahedral mesh and electric field distribution of a 9 R-TML at a frequency of 10~GHz filled with air.}
 }\label{fig:Fig1}
\end{figure}

The basic structure in the numerical calculations is based on a 50 $\Omega$   coaxial transmission line cell for material measurements used in previous studies
(see \cite{Wagn07, Kupf07, Wagn07a}) with an outer diameter of the inner conductor $d_i$ = 7~mm, an inner diameter
of the outer conductor $d_o$ = 16~mm and a length $l$ = 100~mm. Thus, the cutoff frequency of the lowest-order non-TEM
mode $TE_{11}$ is approximately 8~GHz. To study the effects of rod based transmission line configurations in a first step
the diameter of the inner conductor is fixed and the outer conductor is replaced by copper rods with 3.5~mm
diameter (see Fig. \ref{fig:Fig1}) and numbers: 8, 6, 4, 2, 0. The distance between surfaces of inner conductor and rods
is identical to the distance between inner and outer conductor of the C-TML.
Furthermore, the inner conductor was replaced by a 2~mm rod. Thus, the $TE_{11}$ cutoff frequency of the
C-TML increases to approximately 10~GHz. In the second step an experimental setup is developed for
the five R-TML configuration, numerically characterized and used for a comparison of theoretical and experimental obtained results.

Ansoft HFSS solves Maxwell's equations using a finite element method, in which the solution domain is divided into tetrahedral mesh elements (see Fig. \ref{fig:Fig1}). Tangential vector basis functions interpolate field values from both nodal values at vertices and on edges (see \cite{Ansoft2010}). Outer surfaces of the structure between inner and outer conductor or rods in parallel to the x-y plane are used as wave ports (driven modal solution, one mode). A finite conductivity boundary is used at the outer surface normal to the x-y plane of the coaxial cell sample cylinder. The outer surface of the rod based sample cylinder is a radiation boundary and the second-order radiation boundary condition is used. The mesh generation is performed automatically with l/3 wavelength based adaptive mesh refinement at a solution frequency of 10~GHz. Broadband complex S-parameters are calculated with an interpolating sweep (1~MHz - 10~GHz) with extrapolation to DC. The computed generalized S-matrix is normalized to 50 $\Omega$  for comparison of the numerical calculated with measured S-parameters.

\begin{figure}[t]
 \center
  \includegraphics[scale=0.47]{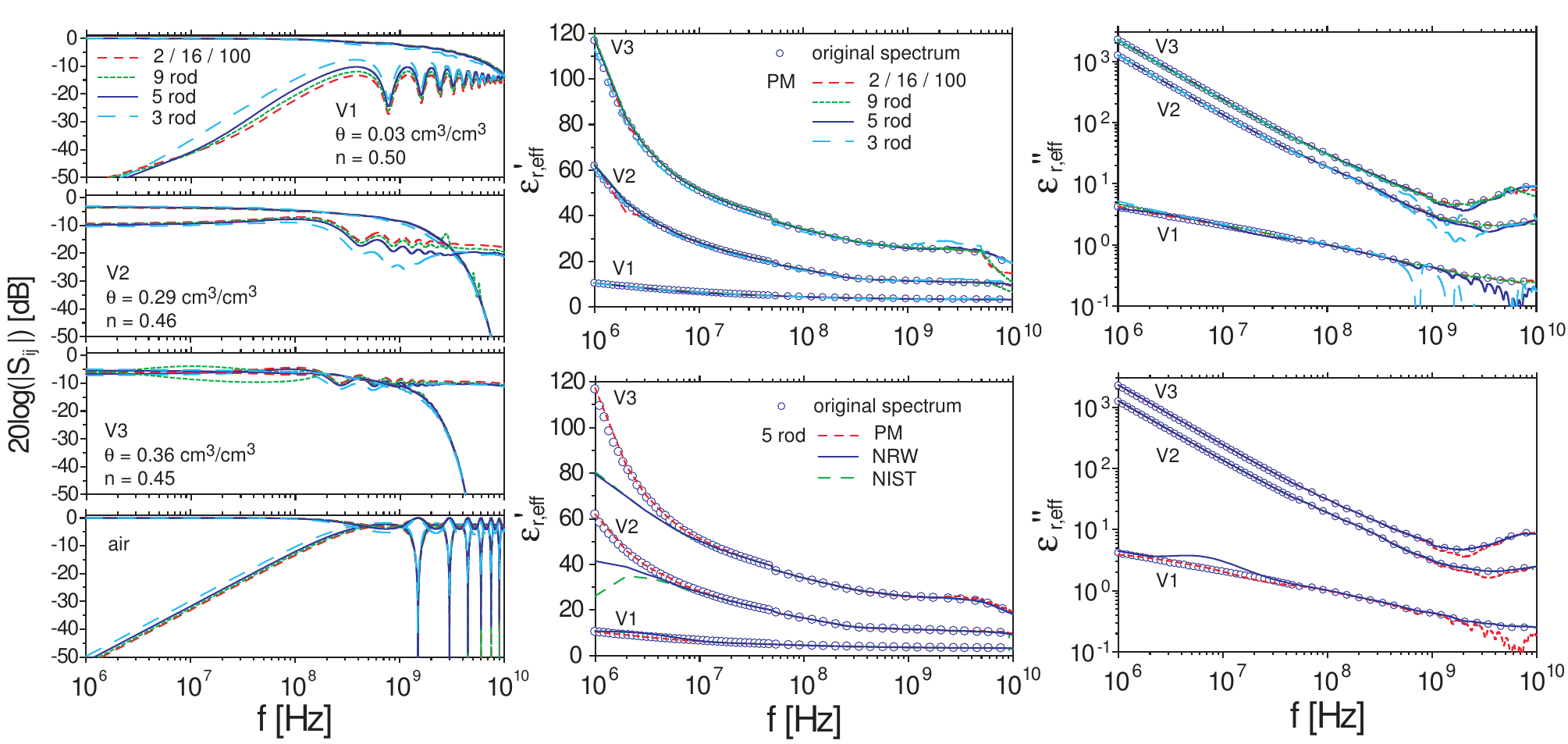}\\
  \caption{\textit{(from left to right) Magnitude of the simulated S-parameters $S_{11}$ and $S_{21}$ of a silty clay loam from a levee at river Unstrut, Germany for three volumetric water contents (V1, V2, V3) in comparison to air. Real part $\varepsilon'_{r,\mbox{eff}}$ and imaginary part $\varepsilon''_{r,\mbox{eff}}$ of the expected and determined complex relative effective permittivity $\varepsilon_{r,\mbox{eff}}^\star$ from numerical results of (top) the 2/16/100 C-TML in comparison to the R-TMLs obtained with PM and (bottom) the five R-TML obtained with PM, NRW and NIST.}
 }\label{fig:Fig2}
\end{figure}

The results of the numerical calculations clearly indicate the disadvantages of R-TMLs in comparison to C-TMLs, i.e. electric and magnetic field distributions as well as radiation, which in addition changes with appropriate electromagnetic material properties (for details see \cite{Wagner2010}). Thus, the transmission line impedance of the empty 2/16/100 mm C-TML as well as R-TMLs exceeds the 50 $\Omega$ impedance of the 7/16/100 C-TML due to the number, diameter and geometrical arrangement of the rods (c.f. Fig. \ref{fig:Fig1}, c). However, 50 $\Omega$ (match) of the R-TML is achieved at a permittivity $>$ 1. Hence, the bandwidth increases for an accurate determination of soil permittivity (Fig. \ref{fig:Fig2}). This is in principle a clear advantage of the R-TMLs compared to the classical 7/16/100 C-TML.

\paragraph{Electromagnetic material properties:} The complex effective permittivity $ \varepsilon_{r,\mbox{eff}}^\star$
of the standard materials and soil used in the numerical calculations were examined in the frequency range from 1~MHz to 10~GHz
at room temperature and under atmospheric conditions with vector network analyzer technique (Rohde \& Schwarz ZVR - DC, 10 kHz
to 4 GHz and PNA E8363B - 10~MHz to 40 GHz). This was performed using a combination of open-ended coaxial-line technique
for liquids (dielectric probe kid HP85070B, high temperature and performance probe, electronic calibration kit N4691B)
and coaxial transmission line technique with mechanical calibration kit and Agilent 85071 software, matlab based algorithms
as well as fitting of all four S-parameters simultaneously under consideration of a dielectric relaxation model
(see \cite{Wagn07}, \cite{Wagn07a} and \cite{Wagner2011} for details).

\paragraph{Inversion procedure:} In general, assuming propagation in TEM mode and non
magnetic materials  $\varepsilon_{r,\mbox{eff}}^\star$ of a sample in a transmission line is related to its
complex impedance $Z_S^\star$  or complex propagation factor $\gamma_S^\star$
as follows (see \cite{Nico70, BJ2004, Gorriti2005b, Wagner2010}):
\begin{eqnarray}
   \varepsilon_{r,\mbox{eff}}^\star & = & \left(\frac{Z_0}{Z^\star_S}\right)^2\label{eq:IM},\\
   \varepsilon_{r,\mbox{eff}}^\star & = & \left(\frac{c_0\gamma^\star_S}{j\omega}\right)^2\label{eq:PMM},\\
   \varepsilon_{r,\mbox{eff}}^\star & = & \frac{c_0 Z_0}{j\omega}\left(\frac{\gamma^\star_S}{Z^\star_S}\right) \label{eq:PM}.
\end{eqnarray}
Herein $Z_0$ is the characteristic impedance of the empty transmission line, $c_0  = \left( {\varepsilon _0 \mu _0 } \right)^{ - 0.5}$  the velocity of light with $\varepsilon _0$ absolute dielectric permittivity, $\mu_0$ absolute magnetic permeability of vacuum, $\omega=2\pi f$ angular frequency and imaginary unit $j=\sqrt{-1}$ . To obtain $Z_S^\star$ or $\gamma_S^\star$ from measured complex S-parameters $S_{ij}$ several quasi analytical approaches are available which lead to either equation (\ref{eq:IM}), (\ref{eq:PMM}) or (\ref{eq:PM}) as shown in \cite{Gorriti2005b}. The numerical calculated complex S-parameters with HFSS were used to compute $\varepsilon_{r,\mbox{eff}}^\star$ in the frequency range between 1~MHz to 10~GHz with the following quasi-analytical methods: classical Nicolson-Ross-Weir algorithm (NRW, \cite{Nico70, Weir74}), a modified NRW algorithm suggested by Baker-Jarvis et al. 2004 (see \cite{BJ2004} section 7.5.4, hereafter called NIST)
and the propagation matrix algorithm (PM, \cite{Gorriti2005a}).\\

The used NRW algorithm utilizes separately two sets of S-parameters $S_{11}$ and $S_{21}$ or $S_{22}$ and $S_{12}$ to obtain a mean
complex effective permittivity. In principle, the three approaches can be used. However, for low loss materials,
the NRW solution is divergent at integral multiples of one-half wavelength in the sample.
The NIST algorithm requires the complete set of S-parameters and is restricted to equation (\ref{eq:PMM}).
For the case of high loss materials both inversion procedure based on (\ref{eq:PMM}) fail in the frequency range
below 100 MHz. The PM algorithm exhibits similar problems at integral multiples of one-half wavelength used with (\ref{eq:IM}) or (\ref{eq:PMM}) as well as  instabilities and failure in the frequency range above 1 GHz used with equation (\ref{eq:PM}).
However, in the frequency range below 100 MHz the determined permittivity with the combined approach (\ref{eq:PM}) is accurate for low as well as high electrical and dielectric loss materials.
Hence, it seems reasonable the results of PM (below  200 MHz with (\ref{eq:PM})) and NRW, NIST or PM (above 100 MHz with (\ref{eq:PMM})) to combine (see in addition \cite{Lauer2010}).  In Fig. \ref{fig:Fig2} the numerical results of the 2/16/100 C-TML in comparison to the R-TMLs are represented for a dispersive soil (for soil details see Wagner et al. 2011 \cite{Wagner2011})
at three volumetric water contents. It is obvious, that the three R-TML is inappropriate for acquisition of broadband
dielectric spectra with the used inversion technique. The most favorable configuration compared to the C-TML offers
the nine R-TML. From a practical viewpoint of minimal disturbance of the sample the five R-TML is a reasonable alternative.

\begin{figure}[p]
 \center
  \includegraphics[scale=0.38]{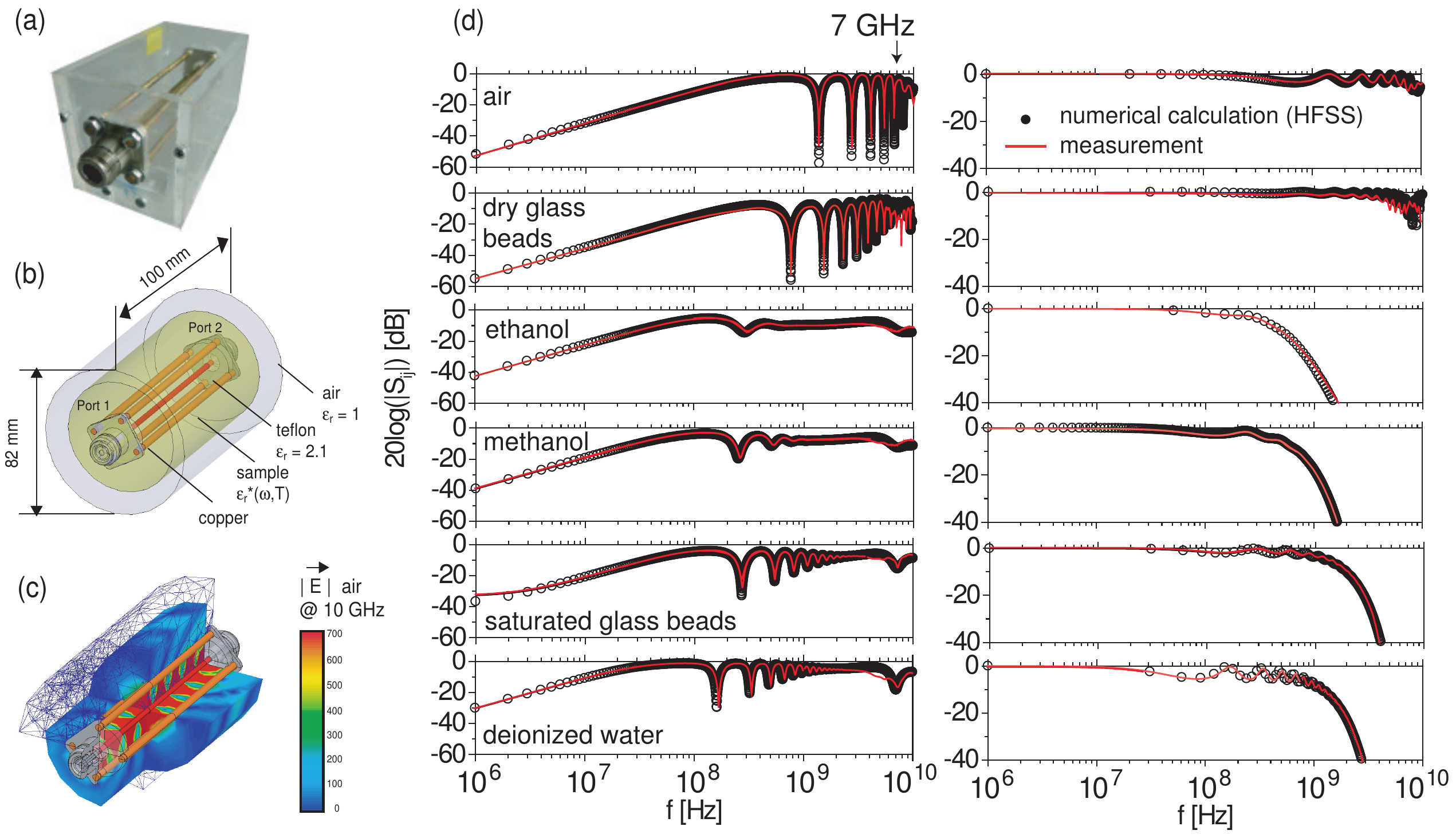}\\
  \caption{\textit{(a) Two port five rod measurement cell, (b) appropriate 3D finite element model,
  (c) electric field distribution (steady state standing wave) in air at 10~GHz. (d)
  Experimental and numerical results in low loss and high loss strong dispersive standard materials
  with magnitude of the complex S-parameters $S_{ij}$ (left) reflection factor $S_{11}$ and
  (rigth) transmission factor $S_{21}$.}}\label{fig:Fig3}
  \vspace{0.3cm}
 \center
  \includegraphics[scale=0.58]{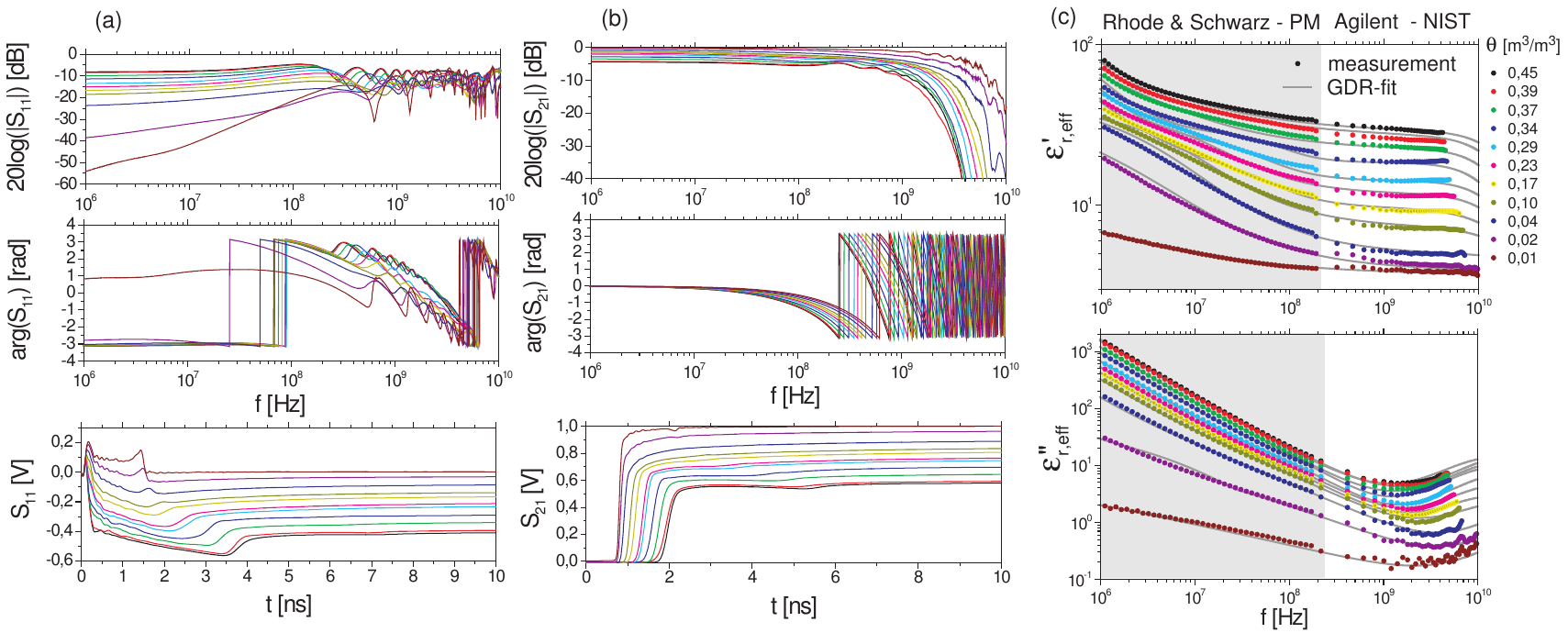}\\
  \caption{\textit{Magnitude, phase shift and TDR-waveform of (a) reflection factor $S_{11}$ and (b) transmission factor $S_{21}$ and (c)
            obtained complex effective relative permittivity $\varepsilon_{r,\mbox{eff}}^\star$
            of the investigated soil for several volumetric water contents $\theta$.}}\label{fig:Fig4}
\end{figure}

\subsection{{Experimental investigations}}

To validate the numerical results a five R-TML was developed (Fig. \ref{fig:Fig3}, a). Fig. \ref{fig:Fig3} (b) shows
the 3D FEM structure for the five R-TML cell and Fig. \ref{fig:Fig3} (c) the magnitude of the electrical field distribution in air at 10~GHz.
Clearly obvious are radiation effects in the frequency range around 10~GHz for low loss materials.
The comparison between numerical and experimental results
in frequency domain are illustrated in Fig. \ref{fig:Fig3} (d) which indicate the consistency of the numerical calculations.
However, in the frequency range above 7 GHz serious differences
between numerical calculations and measurements can be noticed. Appropriate inadequacies between the experimental setup and the model structure are the main reasons which become obvious especially at high frequencies.
Nevertheless, the results indicate that the investigated R-TMLs provide the possibility to obtain accurate high resolution broadband permittivity spectra for high loss materials in the frequency range from 1~MHz to 5~GHz and low loss materials
from 1~MHz to at least 10~GHz.

A slightly plastic clay soil (for soil details see \cite{Wagner2010}) was investigated. The measurements were performed with a
Rhode and Schwarz ZVR (1~MHz - 4 GHz) as well as an Agilent PNA (10~MHz - 10~GHz).
The soil sample is saturated with deionized water, prepared at liquid limit with gravimetric
water content w = 0.267 g/g and placed in the five R-TML (volumetric water content  $\theta$ = 0.41 m$^3$/m$^3$ at a porosity n = 0.41).
Then the sample is stepwise dried isothermal at 23 $^\circ$C under
atmospheric conditions and equilibrated. Appropriate mass loss and sample volume change
were estimated during the drying process to obtain the appropriate volumetric water content.
The frequency dependent complex permittivity was determined as explained by
means of analytical (PM with (\ref{eq:PM}) for f $<$ 200~MHz and NIST with (\ref{eq:PMM}) for f $>$ 200~MHz)
or numerical inversion of measured four complex S-parameters (Fig. \ref{fig:Fig4}).
\begin{figure}[h]
 \center
  \includegraphics[scale=0.55]{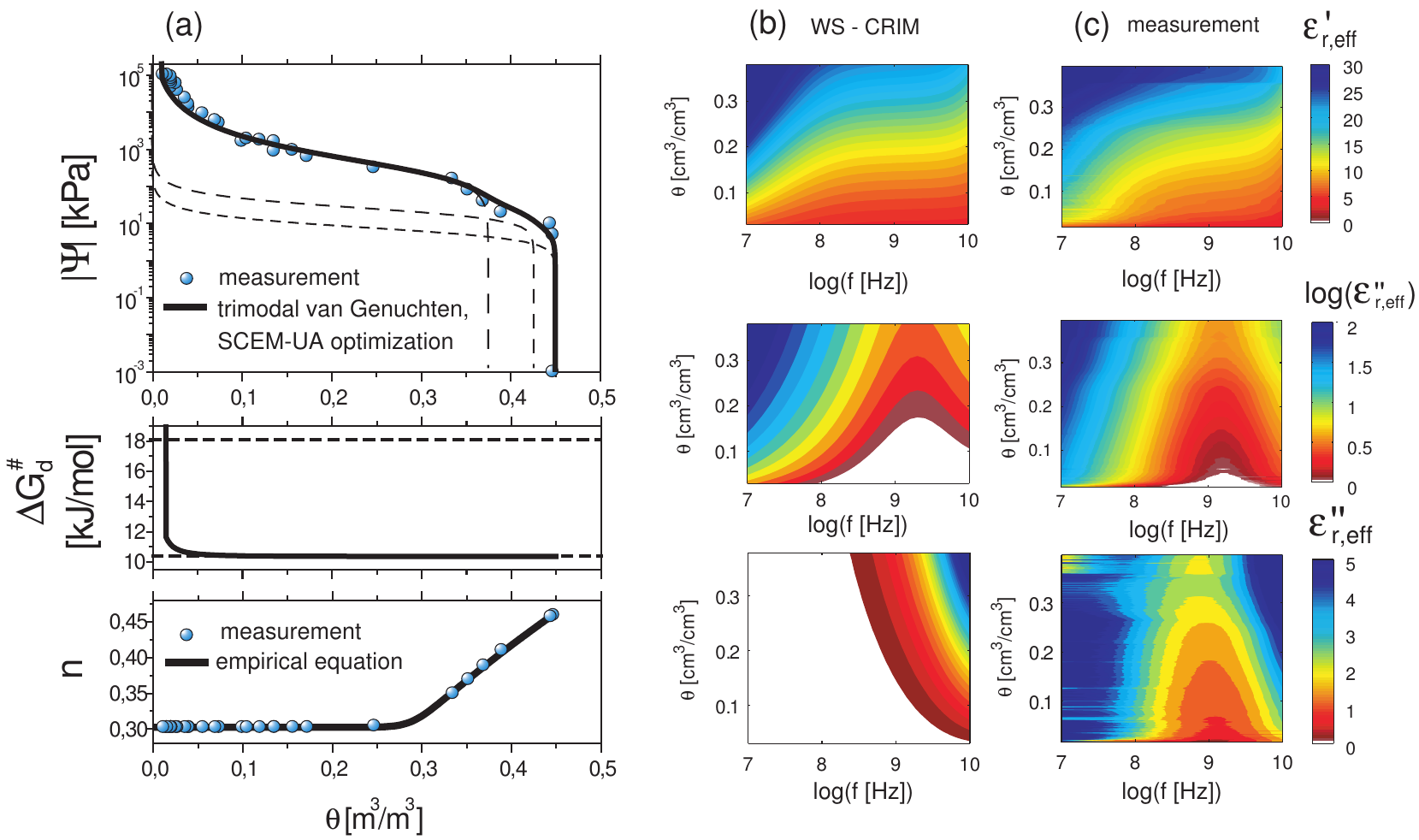}\\
  \caption{\textit{(a, from top to bottom) Matric potential $\Psi$, calculated free enthalpy of activation $\Delta G_d^{\ddag}$ and porosity $n$ as a function of volumetric water content $\theta$. (b, c) Dielectric spectra with (top) real $\varepsilon_{r,\mbox{eff}}'$  and (middle) imaginary $\varepsilon_{r,\mbox{eff}}''$ part of relative effective complex permittivity $\varepsilon_{r,\mbox{eff}}^\star$ as a function of volumetric water content $\theta$ and frequency $f$. (bottom) Dielectric finger print: contributions due to a direct current conductivity are subtracted from the imaginary part, (b) calculated spectra as well as (c) measured.}}\label{fig:Fig5}
\end{figure}
\begin{figure}[t]
 \center
  \includegraphics[scale=0.55]{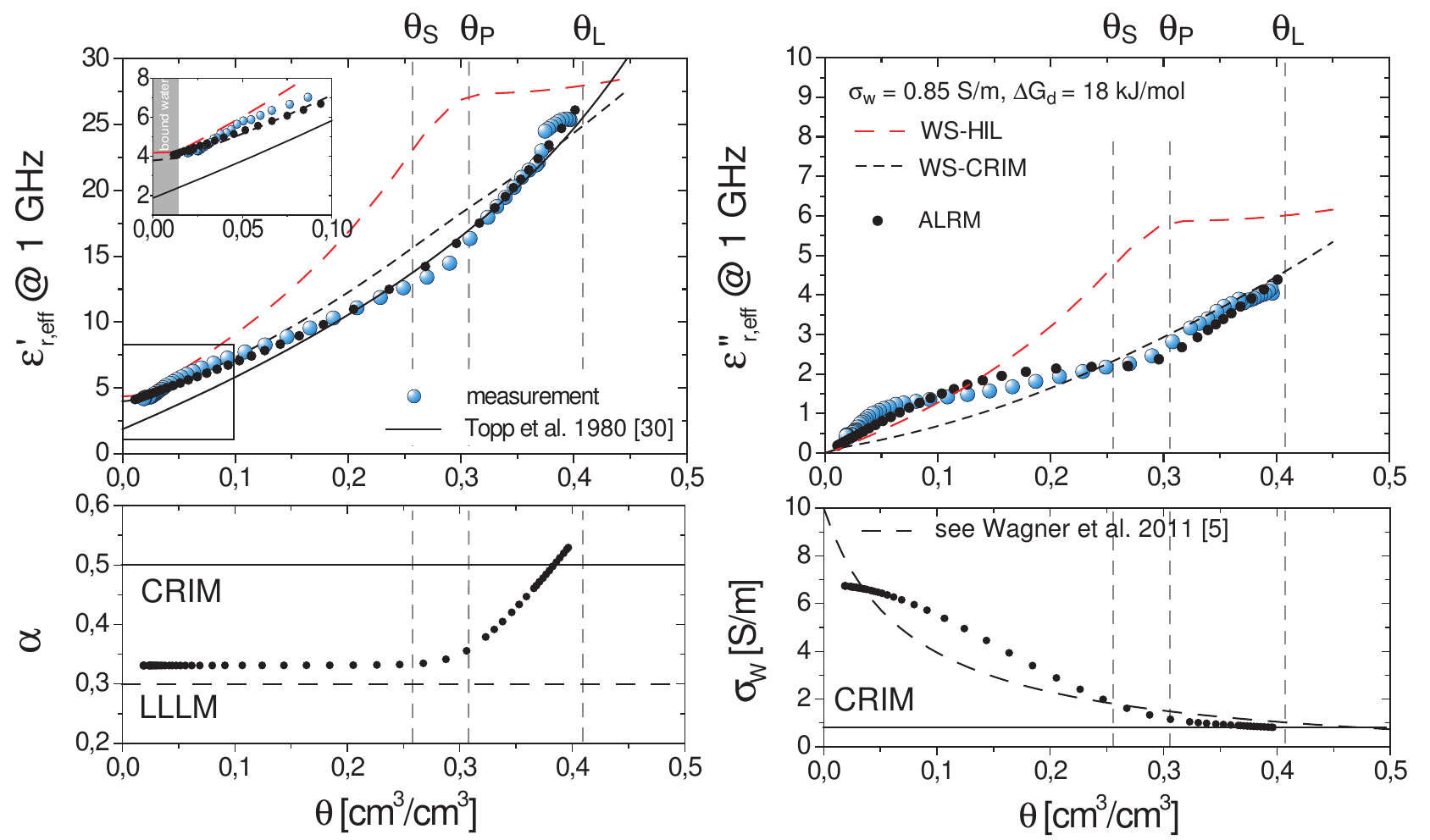}\\
  \caption{\textit{(left, top) Real part $\varepsilon_{r,\mbox{eff}}'$  and (right, top) imaginary part $\varepsilon_{r,\mbox{eff}}''$ of complex effective permittivity $\varepsilon_{r,\mbox{eff}}^\star$ at a measurement frequency of 1 GHz as a function of volumetric water content $\theta$ in comparison to the empirical calibration function according to Topp et al. 1980 [20], WS-HIL, WS-CRIM and ALRM (see text for details). Volumetric water content at the shrinkage limit $\theta_S$, plastic limit $\theta_P$ and at the liquid limit $\theta_L$ are indicated. (left, bottom) Structure parameter $a$ and (right, bottom) effective pore water conductivity $\sigma_{W}$ as a function of $\theta$}.}\label{fig:Fig6}
\end{figure}

\section{\normalsize{DISCUSSION}}

The experimental results are compared with the well known empirical calibration function according to Topp et al. 1980 \cite{Topp80} and the advanced Lichtenecker and Rother model (ALRM) according to equation (\ref{eq:LRMix}) to (\ref{eq:LichtRothM3}). Furthermore the theoretical mixture rule according to equation (\ref{eq:HilMix}) as well as (\ref{eq:LRMix}) in CRIM form are used in four phase form with separately experimental determined soil water characteristic curve (SWCC) as well as shrinkage behavior (see \cite{Schwing2010}, Fig. \ref{fig:Fig5}, a) with the improvements suggested in section \ref{sec:model} (WS-HIL, WS-CRIM). For this purpose, SWCC is parameterized with a trimodal van Genuchten equation according to Priesack and Durner 2011 \cite{Priesack2006} using a shuffled
complex evolution metropolis algorithm (SCEM-UA, \cite{Vrugt2003}) to calculate complex relative dielectric permittivity of free and interface water according to (\ref{eq:matric2}).
Fig. \ref{fig:Fig5} (b) and (c) illustrate the results of the numerical computation according to the modified Hilhorst approach (WS-CRIM) in comparison to experimental determined dielectric spectra. In Fig. \ref{fig:Fig6} calculated and experimental determined complex permittivity $\varepsilon_{r,\mbox{eff}}^\star$ at 1 GHz is represented for WS-CRIM, WS-HIL, ALRM and Topp. The models (WS-HIL, WS-CRIM, ALRM) predict the relative high permittivity at very low water content whereas the Topp-equation clearly underestimate the permittivity. The frequency and water content dependent complex effective permittivity is poorly predicted with WS-HIL while the qualitative characteristics are similar. Substantially better results give WS-CRIM. The deviation between WS-CRIM and experimental obtained imaginary part of effective complex permittivity $\varepsilon_{r,\mbox{eff}}^\star$ especially below the shrinkage limit $\theta_S$ indicates the dependence of the so called structure exponent as well as pore water conductivity on porosity and saturation pointed out by Wagner et al. 2011 \cite{Wagner2011} and considered with ALRM. The relative strong predicted real and imaginary part due to WS-HIL for volumetric water contents above approximately 0.05 m$^3$/m$^3$ is a result of the influence of the porosity in the mixture approach pointed out by Wagner and Scheuermann 2009 \cite{WagScheu2009a}.

\section{\normalsize{CONCLUSION}}

Two-port rod based transmission lines (R-TMLs) were characterized in the frequency range from 1~MHz to 10~GHz and compared to coaxial transmission lines (C-TMLs) by combined theoretical, numerical, and experimental investigations. The propagation characteristics and sensitivity of the R-TMLs were determined by numerical calculations and measurements on standard materials and nearly saturated and unsaturated soils. The results of the numerical calculations clearly indicate the disadvantage of R-TML in comparison to C-TML, i.e. electrical and magnetic field distributions as well as radiation, which in addition changes with appropriate investigated electromagnetic material properties. However, 50 $\Omega$ (match) of the R-TMLs are achieved at a permittivity $>$ 1. Thus, the bandwidth increases for an accurate determination of the permittivity. This is in principle a clear advantage of the R-TMLs compared to the classical C-TML. To validate the numerical calculations a five R-TML was developed. The numerical results are in reasonable agreement with appropriate experimental measurements. Hence, R-TMLs provide the possibility to obtain accurate high resolution broadband permittivity spectra for high loss materials in the frequency range from 1~MHz to 5 GHz and low loss materials
from 1~MHz to at least 10~GHz. To analyze the coupled hydraulic and dielectric soil properties a slightly plastic clay soil was investigated. The results indicate that the bound water contribution of the examined soil is substantially lower than expected. Furthermore, the influence of the pore water conductivity and soil structure has to taken into account in the modelling approach for coupling hydraulic and dielectric soil properties.

\section*{\normalsize{ACKNOWLEDGEMENTS}}

The authors grateful acknowledge the German Research Foundation (DFG) for support of the projects Wa 2112/2-1 and SCHE 1604/2-1.

\small
\bibliographystyle{ieeetr}
\bibliography{Literatur1}

\end{document}